%% file: root.tex
\documentclass[letterpaper, 10 pt, conference]{ieeeconf}  

\IEEEoverridecommandlockouts                              

\usepackage{graphics} 
\usepackage{amsmath} 
\usepackage{amssymb}  

\usepackage{flushend}  

\usepackage{cite}
\usepackage{graphicx}
\usepackage{color}

\newcommand{\be}{\begin{enumerate}}
\newcommand{\ee}{\end{enumerate}}
\newcommand{\Tr}{\text{Tr}}

\newcommand{\Exp}{\mathbb E}

\newtheorem{prop}{Proposition}
\title{\LARGE \bf
Localizability-Constrained Deployment of Mobile Robotic Networks with Noisy Range Measurements
}

\author{Jerome Le Ny$^*$ and Simon Chauvi\`ere
\thanks{$^*$Corresponding author. This work was done while the second author 
was visiting Polytechnique Montreal.}
\thanks{J. Le Ny is with the Department of Electrical Engineering, 
Polytechnique Montreal, and with GERAD, Montreal, QC H3T 1J4, Canada. 
	{\tt\small jerome.le-ny@polymtl.ca}}%
\thanks{S. Chauvi\`ere is with the \'Ecole Normale Sup\'erieure de Cachan,
94230 Cachan, France.
	{\tt\small simon.chauviere@ens-cachan.fr}}%
}

\begin{document}

\maketitle
\thispagestyle{empty}
\pagestyle{empty}

\begin{abstract}
%
%
When nodes in a mobile network use relative noisy measurements with respect to their neighbors to estimate their positions, 
the overall connectivity and geometry of the measurement network has a critical influence on the achievable localization accuracy.
This paper considers the problem of deploying a mobile robotic network implementing a cooperative localization scheme based
on range measurements only, while attempting to maintain a network geometry that is favorable to estimating the robots' positions
with high accuracy. 
The quality of the network geometry is measured by a ``localizability'' function serving as potential field for robot motion planning.
This function is built from the Cram\'er-Rao bound, which provides for a given geometry a lower bound on the covariance matrix 
achievable by any unbiased position estimator that the robots might implement using their relative measurements. 
We describe gradient descent-based motion planners for the robots that attempt to optimize or constrain
different variations of the network's localizability function, and discuss ways of implementing these controllers 
in a distributed manner.
Finally, the paper also establishes formal connections between our statistical point of view and maintaining a form of weighted rigidity
for the graph capturing the relative range measurements.
\end{abstract}


\input{intro.tex}

\input{statement.tex}

\input{centralized.tex}
\input{distributed.tex}

\input{simulations.tex}

\section{Conclusions}

We have considered the problem of deploying a mobile robotic network implementing 
a cooperative localization scheme to estimate the robots' positions from noisy relative 
range measurements, by restricting the trajectories of the robots to maintain group
shapes that enable sufficiently accurate position estimation.
Concretely, a potential-field based motion planner aims to maintain network 
configurations such that the Cram\'er-Rao Lower Bound on the variance of
any unbiased position estimator constructed based on the distance measurements is small.
We established connections between this methodology and the problem of rigidity
maintenance control, by remarking that the Fischer information matrix (FIM) can be viewed
as a type of weighted rigidity matrix. We also discussed distributed implementations of
the gradient descent motion planner, for different types of potentials built from the FIM.


\bibliographystyle{IEEEtran}
\bibliography{IEEEabrv,biblio}

\end{document}

%% file: intro.tex
\section{Introduction}


Deploying an unmanned vehicle system to carry out a mission in a given environment 
depends crucially on our ability to precisely localize each vehicle in that environment 
in real-time.
In some cases, it is not feasible to equip each vehicle with an absolute positioning system
such as a GPS receiver, e.g., because of lack of GPS signal availability 
due to jamming or when deploying the vehicles indoors, underwater or in covered areas.
In such cases, one alternative is to use relative measurements between the robots,
i.e., relative distance and/or bearing measurements that can be obtained by
on-board equipment such as lasers, cameras or RF transceivers,
and to let the robots estimate their positions ``cooperatively'' based on these
relative measurements. A small number of robots or beacons with known positions 
in a global common reference frame can then be sufficient to localize each robot in that frame.

Many cooperative localization algorithms have been proposed to estimate the 
positions of nodes from relative measurements between them, with applications 
to sensor networks and multi-robot systems among others, see, e.g., 
\cite{savvides:Mobicom2001:cooperativeLoc, moore:conf2004:robustLoc, cao2006sensor, sheu2010distributed}.
In this paper, we focus on scenarios where the robots can only measure the relative 
range of their neighbors (rather than making full relative position measurements), 
which is motivated by the availability of 
low cost sensors capable of performing such measurements accurately 
with low power and computational requirements, such as short-range 
Ultra-Wide Band (UWB) transceivers measuring the time of flight (ToF) 
of signals exchanged between the robots \cite{Gezici:SPM05:UWBlocalization}.


Compared to static sensor networks, mobile robotic networks offer the possibility
of controlling the motion of the robots to assist the cooperative localization system.
Indeed, it is well-known that the geometric configuration of the ranging network, i.e., the
network with links defined by the pairs of robots making relative measurements, can have 
a drastic impact on the positioning accuracy achievable by cooperative localization 
algorithms \cite{patwari2005locating}, and hence the allowed configurations should
be constrained during deployment of the robots if such a positioning scheme is used. 
This coupling between motion planning and ``localizability'', i.e., the ability to estimate 
the robots' positions under a cooperative localization scheme, has received much less 
attention than the development of actual localization algorithms. 

One recent approach related to localizability-constrained motion planning, due to 
Zelazo et al. \cite{Zelazo:RSS12:rigidity, zelazo2015decentralized}, aims to maintain 
the \emph{rigidity} of the ranging network as the robots move. 
Intuitively, rigidity \cite{roth:AM1981:rigidFrameworks} is a property 
that the network can exhibit if enough ranging measurements are available 
to constrain its shape, except perhaps for some large deformations such as 
symmetric rearrangements of subsets of nodes. This property is known to 
be tightly linked to the possibility of uniquely reconstructing the nodes' positions 
from range measurements \cite{eren2004rigidity}, and hence is useful
to enforce for a robotic network. However, it is essentially a combinatorial
property that does not take into account range measurement errors and their 
statistical characteristics, and in particular it is insufficient by itself
to predict the achievable accuracy of the position estimates under 
noisy range measurements.
For example, a set of robots that are almost aligned can form a rigid network,
e.g., if each robot measures its distance with respect to all others, yet it will be essentially
impossible for them to estimate their positions accurately as soon as small errors
contaminate the range measurements, these errors being greatly amplified by the poor 
network geometry.

The position estimation performance achievable for a given geometry under noisy relative measurements 
is more accurately captured by statistical notions such as the Cram\'er-Rao Bound (CRB) \cite{VanTrees:Book69:SP}, 
which provides a lower bound on the covariance matrix of any possible unbiased position estimate.
Several references compute the CRB for sensor networks performing cooperative
localization with a variety of sensing modalities
\cite{stoica1998cramer, taylor2003computing, patwari2005locating, mohammadi2012decentralized},
but with a focus on static nodes.
The CRB has also been used extensively to guide the motion of mobile sensors 
performing signal processing tasks such as tracking a target \cite{hernandez2004multisensor} 
or estimating the parameters of an spatial process \cite{Ucinski:book05:optimalMeas}.
In contrast, we use the CRB as the basis for a multi-robot motion planning strategy that supports
more accurate position estimates for the robots themselves.

We formulate our problem more precisely in Section \ref{section: pb statement},
where we also show that there is in fact a close connection between the CRB
and a certain notion of \emph{weighted} rigidity introduced 
in \cite{Zelazo:RSS12:rigidity, zelazo2015decentralized}, for a specific choice of weights.
In Section \ref{section: potential fields}, we propose a potential-field based motion
planning method to guide the robots along trajectories that maintain the CRB of the network
low, as a measure of the localizability of the robots. Descending the gradient of
a potential field is a standard tool to deploy groups of robots performing various
tasks, from source seeking to formation control 
and coverage control \cite{Wang91_navigation, Reif99_socialPotentials, Bullo09_book, 
krick2009stabilisation, LeNy:TAC2013:adaptiveDeployment}. 
Potential fields can encode constraints on the robots' paths such as obstacle avoidance \cite{Khatib86_potential},
communication constraints \cite{LeNy:JSAC12:adaptiveComm} 
and, as we discuss here, localizability constraints.
%
%
It is generally desirable with such methods to obtain motion planing algorithms that can be implemented 
in a distributed fashion, with the robots communicating only with a restricted number of neighbors, but
possibly 
through multiple iterations.
This issue is addressed in Section \ref{section: distributed algos}.
Finally, Section \ref{section: simu} summarizes our approach and briefly illustrates 
it via simulations. 

%% file: statement.tex

\section{Problem Statement and \\Localizability Definition}
\label{section: pb statement}


Consider a network of $n+m$ mobile robots evolving in a $2D$ space, i.e., with positions 
$\mathbf p_i = [x_i,y_i]^T \in \mathbb R^2$, $i=1,\ldots,n+m$ expressed in a global common reference frame. 
We assume that $m$ robots know their position perfectly, 
e.g., they could in fact be static nodes whose position has been carefully determined. We call these $m$ robots \emph{anchors}
and without loss of generality we choose their indices to be $n+1, \ldots, n+m$.
Let $\tilde {\mathbf p} = [\mathbf p_1^T, \ldots, \mathbf p_n^T]^T \in \mathbb R^{2n}$ be a column vector containing the (unknown) positions 
of the remaining robots, and let $\mathbf p =  [\mathbf p_1^T, \ldots, \mathbf p_{n+m}^T]^T \in \mathbb R^{2(n+m)}$.

Any robot $i$, including the anchors, can measure with some noise its Euclidean distances $d_{ij} = \|\mathbf p_i - \mathbf p_j\|$ with respect to 
other robots $j$ in a set $\mathcal N_i \subset \{1,\ldots,n+m\}$, which we call its \emph{neighbors}, and moreover it can 
also communicate with these neighbors (in order to implement a distributed motion planning algorithm). 
For simplicity, we assume in this paper that range measurement capabilities 
are symmetric, i.e., $j \in \mathcal N_i$ if and only if $i \in \mathcal N_j$.
The sets $\mathcal N_i$ could change over time, as the geometry of the network evolves, or be dependent on 
the global configuration $\mathbf p$. 
The agents with their sets of neighbors then form an undirected \emph{ranging graph} 
$\mathcal G = (\mathcal V, \mathcal E)$, with $|\mathcal V| = n+m$ vertices so that 
an edge $\{i,j\} \in \mathcal E$ if $i$ and $j$ are neighbors. 
We also define the indicator function $\mathbf 1_{\mathcal N_i}$ for a set $\mathcal N_i$ by 
$\mathbf 1_{\mathcal N_i}(j) = 1$ if $j \in \mathcal N_i$ and $\mathbf 1_{\mathcal N_i}(j) = 0$ otherwise.


\subsection{Range Measurements and Cooperative Localization}

For two agents $i, j$ capable of measuring their distance $d_{ij}$, we denote 
by $\theta_{ij}$ a measurement by $i$ of its distance to $j$ (which could be different from $\theta_{ji}$,
because of measurement errors).
We consider here only simple ranging measurement models, namely, the case of 
distance measurements with random additive Gaussian errors for all nodes, 
i.e., $\theta_{ij} = d_{ij} + \epsilon_{ij}$, or with random multiplicative log-normal errors
for all nodes, i.e., $\theta_{ij}=e^{\epsilon_{ij}} d_{ij}$ or equivalently $\log(\theta_{ij})=\log(d_{ij})+\epsilon_{ij}$,
with in both cases $\epsilon_{ij} \sim \mathcal{N}(0,\,\sigma^{2})$, and the same parameter $\sigma$ 
for all $i, j$. 
We make the standard simplifying assumption that all measurement errors $\epsilon_{ij}$ are independent.
Additive errors are characteristic of ToF-based distance measurements, and multiplicative
errors of measurements based on signal strength for example \cite{patwari2005locating}. 

Starting from the relative distance measurements $\theta_{ij}$, the robots must estimate their positions in the 
global common frame of reference.
In other words, they must form an estimate $\hat{\mathbf p}$ of $\mathbf p$. 
We assume $m \geq 3$, so that we have enough anchors in order to be able to remove the intrinsic translational 
and rotational ambiguity for the whole network associated with relative range measurements. 
The literature addressing this \emph{cooperative localization problem} is extensive, with both centralized 
and decentralized algorithms available \cite{savvides:Mobicom2001:cooperativeLoc, moore:conf2004:robustLoc, cao2006sensor, sheu2010distributed}.
For example, a basic method could be to solve a least-squares problem minimizing the sum of the squared
residuals $r_{ij}^2 = |\theta_{ij}^2 -\|\hat{\mathbf p}_i - \hat{\mathbf p}_j\|^2|^2$, which leads via gradient descent
to distributed computations of an estimate $\hat{\mathbf p}$.
Our deployment methodology is independent of the choice of cooperative localization algorithm implemented 
by the robots, but requires a real-time estimate $\hat{\mathbf p}$ provided by such an algorithm, as a input signal 
to the motion planner.


\subsection{Lower Bound on Localization Accuracy}
\label{eq: FIM expressions}


The CRB \cite{VanTrees:Book69:SP} provides a lower bound on the covariance matrix of any unbiased 
position estimate $\hat{\mathbf p}$ constructed from the relative measurements $\Theta := [\theta_{ij}]_{i,j \in \mathcal N_i}$ 
between the robots. This bound depends on the relative positions of the robots, i.e., on the geometry of
the ranging network.
While it is not necessarily achieved by a particular localization algorithm, we use it here as an indication 
of the ability of a network geometry to support accurate position estimation, i.e., of the \emph{localizability}
of the network.
\emph{Our objective is to design motion planning strategies for the robot network that maintain a high
level of localizability}, which we aim to achieve concretely by 
maintaining ranging network geometries associated with a small CRB.

For a measurement model given by the measurements' probability density $h(\Theta | \mathbf p)$,
we define the (symmetric) $2n\times2n$ Fischer Information matrix (FIM) $F( \mathbf p)$ by
\[
F(\mathbf p) = -\mathbb E\left[ \frac{\partial^2}{\partial \tilde{\mathbf p}^2}\ln h(\Theta |  \mathbf p) \right],
\]
where the matrix inside the expectation operator is the Hessian matrix of $\ln h(\Theta |  \mathbf p)$
with respects to the (unknown) position variables $x_i, y_i, i=1,\ldots,n$.
The CRB states that 
\[
\text{Cov}(\hat{\mathbf p}) \succeq (F( \mathbf p))^{-1},
\]
for any estimate $\hat{\mathbf p}$ of $\mathbf p$ constructed from the measurements 
$\Theta$ that is unbiased, i.e., such that $\Exp[\hat{\mathbf p}] = \mathbf p$. 
The notation $A \succeq B$ means that $A-B$ is positive semi-definite.



For our measurement models, the FIM can be computed explicitly, see \cite{patwari2005locating} 
for example. With our variable ordering for  $\mathbf p$, it can be written 
in the form $F(\mathbf p) = [F_{ij}(\mathbf p)]_{i,j=1}^n$, where the blocks 
$F_{ij}$ for $i \neq j$ (defined also for indices in $\{n+1, \ldots, n+m\}$), are the $2 \times 2$ matrices
\begin{align} \label{eq: non-diag block FIM}
&F_{ij}(\mathbf p) := F_{ij}(\mathbf p_i, \mathbf p_j) = \\
& - \frac{\mathbf 1_{\mathcal N_i}(j)}{\sigma^2 d_{ij}^{2\alpha}} 
\begin{bmatrix} (x_i - x_j)^2 & (x_i-x_j)(y_i-y_j) \\ (x_i-x_j)(y_i-y_j) & (y_i-y_j)^2 \end{bmatrix}, \nonumber
\end{align}
 
and moreover the diagonal $2 \times 2$ blocks are 
\begin{align} \label{eq: diag block FIM}
&F_{ii}(\mathbf p) = - \sum_{k \in \mathcal N_i} F_{ik}(\mathbf p_i, \mathbf p_k) = \\
& \sum_{k \in \mathcal N_i} \frac{1}{\sigma^2 d_{ik}^{2\alpha}}  
\begin{bmatrix} (x_i - x_k)^2 & (x_i-x_k)(y_i-y_k) \\ (x_i-x_k)(y_i-y_k) & (y_i-y_k)^2 \end{bmatrix}. \nonumber
\end{align}

In the expressions \eqref{eq: non-diag block FIM} and \eqref{eq: diag block FIM}, we set 
$\alpha=1$ for additive Gaussian noise and $\alpha=2$ for multiplicative log-normal noise. 
Note that the matrix $F$ is indeed symmetric since $F_{ij}(\mathbf p_i, \mathbf p_j) = F_{ji}(\mathbf p_j, \mathbf p_i)$, 
with moreover each block $F_{ij}$ and $F_{ii}$ also symmetric.
In addition, the sparsity pattern of $F$ is in correspondance with the links in the ranging graph $\mathcal G$, i.e.,
$F_{ij} = 0$ if $j \notin \mathcal N_i$.
Note also that the matrices $F_{ij}$ for $j \neq i$ present in $F$ only depend on the unknown position variables 
$\tilde{\mathbf p}$, however the diagonal matrices $F_{ii}$ can involve the anchor variables $\mathbf p_{n+1}, \ldots \mathbf p_{n+m}$,
since the anchors are included in some of the sets $\mathcal N_i$. This fact is important to be able to obtain a FIM that
is invertible.
%
%

\subsection{Connections with Rigidity Theory}
\label{section: rigidity connection}

We conclude this section by making a connection between the FIM above and weighted rigidity theory \cite{zelazo2015decentralized},
and show that up to reordering of the variables, the FIM can be viewed as (a submatrix of) a weighted 
Laplacian matrix \cite{godsil:2013book:AGT} of the ranging graph $\mathcal G$. 
First, define $\bar F$ to be the $2(n+m) \times 2(n+m)$ matrix with 
blocks $F_{ij}$ as for $F$, but including the blocks corresponding to the anchor nodes.
Next, reorder the coordinate variables from the order defining $\mathbf p$ to $[x_1, \ldots, x_n, y_1, \ldots, y_n]$
and let $P$ denote the permutation matrix describing this change of coordinates.
The (extended) FIM, built from the entries of \eqref{eq: non-diag block FIM}, \eqref{eq: diag block FIM}
with this new ordering, denoted $\bar F_1 := P \bar F P^{-1}$, is of the form
\[
\bar F_1 = \begin{bmatrix}
\bar  F_{xx} & \bar F_{xy} \\ \bar F_{xy}^T & \bar F_{yy}
\end{bmatrix}.
\]
Now, orient the ranging graph arbitrarily and define the following $|\mathcal E| \times 2(n+m)$ \emph{weighted rigidity matrix} $R(\mathbf p)$ 
as the matrix with one row per edge in $\mathcal E$, and such that if $(i,j) \in \mathcal E$, the corresponding row is
\[
\begin{bmatrix}
\mathbf 0^T & \frac{x_i-x_j}{\sigma d_{ij}^\alpha} & \mathbf 0^T & - \frac{x_i-x_j}{\sigma d_{ij}^\alpha} &
\mathbf 0^T & \frac{y_i-y_j}{\sigma d_{ij}^\alpha} & \mathbf 0^T & - \frac{y_i-y_j}{\sigma d_{ij}^\alpha}
\end{bmatrix},
\]
where $\mathbf 0^T$ denotes a zero row vector of appropriate dimensions, and the non-zero entries
are in the columns $i$, $j$, $n+m+i$ and $n+m+j$. Straightforward calculations lead to the following
result.

\begin{prop}
We have $\bar F_1(\mathbf p) = R^T(\mathbf p) R(\mathbf p)$.
\end{prop}

From this remark, the following proposition follows from \cite[Proposition 2.15]{zelazo2015decentralized}.
Define $Q_{\mu \nu}(\mathbf p)$ for $\mu$ and $\nu$ equal to $x$ or $y$ as the $|\mathcal E| \times |\mathcal E|$ 
diagonal matrix with entry $\frac{(\mu_i-\mu_j)(\nu_i-\nu_j)}{\sigma^2 d_{ij}^{2\alpha}}$ for edge $(i,j) \in \mathcal E$, 
with edges ordered as for the rows of $R$. 
Let $B$ be the incidence matrix of the graph $\mathcal G$, i.e., the $(n+m) \times |\mathcal E|$ matrix with
entries $B_{i,(i,j)} = +1$, $B_{j,(i,j)} = -1$, and zero otherwise.

\begin{prop} \label{prop: rigidity connection}
We have $\bar F_1(\mathbf p) = (I_2 \otimes B) Q(\mathbf p) (I_2 \otimes B^T)$, where
$ \otimes$ denotes the Kronecker product and
\[
Q(\mathbf p) = \begin{bmatrix}
Q_{xx}(\mathbf p) & Q_{xy}(\mathbf p) \\ Q_{yx}(\mathbf p) & Q_{yy}(\mathbf p)
\end{bmatrix}.
\]
\end{prop}

\vspace{0.1cm}
Proposition \ref{prop: rigidity connection} shows that $\bar F_1$ is 
a (weighted) symmetric rigidity matrix as introduced in \cite{Zelazo:RSS12:rigidity, zelazo2015decentralized}, 
for a specific set of weights, namely, edge $(i,j) \in \mathcal E$ has weight $1/(\sigma d_{ij}^\alpha)$. 
As a result, the techniques developed in these papers for rigidity maintenance are applicable to 
keep the CRB low, at least for the E-optimal design approach introduced in the next section.
Note also that our weights diverge as the agents get closer, in contrast to the
weights introduced in \cite{Zelazo:RSS12:rigidity, zelazo2015decentralized} for purposes such as collision
avoidance, which remain bounded.

\section{Potential Field Based Motion Planning}
\label{section: potential fields}

A standard technique to design multi-robot deployment algorithms is to let the robots descend the
gradient of a potential field (cost function) encoding constraints such as collision avoidance
or connectivity and tasks such as coverage-control or source seeking \cite{Bullo09_book}. 
Here we use this methodology to maintaining good localizability for the group.

Given a real-valued potential function $f(\mathbf p)$ measuring the quality of a geometric 
configuration $\mathbf p$, with lower values corresponding to higher quality configurations,
potential field based motion planners design trajectories for the robots by obtaining successive
configurations $\mathbf p^{0}, \mathbf p^{1}, \ldots$ that descend the gradient of $f$, i.e.,
\begin{align}	\label{eq: gradient descent}
\mathbf p^{k+1} = \mathbf p^{k} - \gamma_k \nabla f |_{\mathbf p^{k}},
\end{align}
where $\gamma_k$ are some stepsizes, which could be taken constant. 
For instance, obstacle avoidance controllers can be obtained by designing 
functions that increase sharply in the neighborhood of an obstacle \cite{Khatib86_potential}.
The dynamics of the robots are often neglected at this stage, as we do here, and a lower
level controller is then necessary to track the resulting trajectories with physical platforms.

For $f$ sufficiently smooth, the sequence \eqref{eq: gradient descent} will tend
to configurations that remain in a neighborhood of a local minimum of $f$, and indeed 
most multi-robot potentials can have many such minima.
A further complication comes from the fact that here the current configuration $\mathbf p^k$ 
is not known exactly but is estimated as $\hat{\mathbf p}^k$ from a cooperative localization
algorithm, in which
case one can implement
\begin{align}	\label{eq: gradient descent - estimate}
\mathbf p^{k+1} = \mathbf p^{k} - \gamma_k \nabla f |_{\hat{\mathbf  p}^{k}}.
\end{align}
Errors in the position estimates can lead to errors in the update directions, 
but a formal discussion of this issue is outside of the scope of this paper.

\subsection{Choice of Potential Function}

For illustration purposes, consider potential functions of the form
\begin{equation}	\label{eq: global potential}
f(\mathbf p) = f_{loc}(\mathbf p) + \alpha f_{conn}(\mathbf p) + \beta f_{task}(\mathbf p),
\end{equation}
with $\alpha, \beta$ some parameters weighting particular components of the potential field.
The function $f_{task}$ aims to deploy the robots to achieve a specific task, 
for example reach a specific goal in the workspace, or cover an area \cite{Bullo09_book}. 
Many such potentials have been designed for multi-robot systems. For concreteness,
consider the function 
\begin{equation}	\label{eq: task potential}
f_{task}(\mathbf p) = \frac{1}{2}  \overset{n}{\underset{i=1}{\sum}}(x_i-\bar x_{i})^2. 
\end{equation}
This function drives the robots toward configurations where robot $i$ is close to a desired line of $x$-coordinate 
$\bar x_{i}$.
An important aspect in the choice of a potential function is to facilitate distributed gradient computations.
Indeed, the updates \eqref{eq: gradient descent - estimate} can be rewritten for agent $i$ as
\begin{equation}	\label{eq: individual gradient step}
\mathbf p_i^{k+1} = \mathbf p_i^{k+1} - \gamma_k \frac{\partial f}{\partial \mathbf p_i}(\hat{\mathbf p}^k),
\end{equation}
where $\partial f / \partial \mathbf p_i$ represents the vector $\begin{bmatrix} \partial f / \partial x_i, \partial f / \partial y_i \end{bmatrix}$.
Ideally, computing $\frac{\partial f}{\partial \mathbf p_i}(\hat{\mathbf p}^k)$ for robot $i$ should be possible by communicating
only with a few other robots (its neighbors), to facilitate scaling of the algorithm with the size of the network and avoid 
communication or computation bottlenecks at certain nodes. 
Distributed gradient descent updates are trivial for \eqref{eq: task potential}, involving in fact no communication
between robots, since 
$\partial f_{task}(\hat{\mathbf p}) / \partial \mathbf p_i = [(\hat x_i-\bar x_i), 0]$.

%

We include in \eqref{eq: global potential} a potential $f_{conn}$ to maintain certain pairs of robots
sufficiently close. This imposes a priori that we always want to have range measurements 
during deployment between these pairs.
In fact, the localizability potential discussed next could in principle lead to maintaining
appropriate links during deployment, but in practice this can lead to numerical difficulties if
too many links suddenly disappear, an issue that is left for future work.
%
%
Denote by $\mathcal{E}_{cons}$ the set of links to guarantee. 
Nodes linked by an edge in $\mathcal{E}_{cons}$ can be kept within a distance $d_{\max}$ 
by using a barrier potential such as
\[
f_{conn}(\mathbf p) = \sum_{(i,j)\in \mathcal{E}_{cons}} g(\| \mathbf p_i - \mathbf p_j \|),
\]
where $g(d) = 0$ if $d < d_0$ and
$g(d) = \left(\frac{1}{d_{\max}-d} - \frac{1}{d_{\max}-d_0}\right)^2$
if $d_0 \leq d < d_{\max}$.
Here $d_0$ is the distance at which the inter-agent distance starts to be penalized.
It is straightforward to see that $\frac{\partial f_{conn}}{\partial \mathbf p_i}$ can be computed
by agent $i$ from the position information of its neighbors in $\mathcal E_{cons}$ only.

\subsection{Localizability Potential}

The remaining term $f_{loc}$ in \eqref{eq: global potential} is the main focus of this paper.
We build a potential function from the FIM to attempt to restrict the motion of the group to 
configurations where the CRB, i.e., $(F(\mathbf p))^{-1}$ is sufficiently small. Since the potential
field must be scalar, some information provided by the matrix inequality in the CRB will be lost. 
Various potential functions can be envisioned, as suggested by the literature
on optimal design of experiments in statistics \cite{Pukelsheim93_optimalDesign}. 
For example, one can try to reach configurations that minimize one of the
following functions taken as $f_{loc}$
\begin{align}
f_{T}(\mathbf p) &= - \Tr[F(\mathbf p)] \;\;\; (\text{T-optimal design}) \label{eq: T optimal cost} \\
f_{D}(\mathbf p) &= -\ln \det(F(\mathbf p)) \;\;\; (\text{D-optimal design}) \label{eq: D optimal cost}  \\
f_{A}(\mathbf p) &= \Tr[F(\mathbf p)^{-1}] \;\;\; (\text{A-optimal design}) \label{eq: A optimal cost} \\
\text{or } f_E(\mathbf p) &= - \lambda_{\min}(F(\mathbf p)) \;\;\; (\text{E-optimal design}). 
\label{eq: E optimal cost} 
\end{align}

The function $f_T$ is the easiest to compute and minimize, unfortunately it typically leads to
undesirable paths and configurations, in particular because it does not prevent $F(\mathbf p)$
to become singular.
The function $f_E$, where $\lambda_{\min}$ is the minimum eigenvalue of the FIM, is 
essentially the potential adopted in previous work on connectivity maintenance 
\cite{yang:automatica10:connectivity} (maximizing the first nonzero eigenvalue of the 
Laplacian matrix of the communication graph) and rigidity maintenance \cite{zelazo2015decentralized} 
(maximizing the first nonzero eigenvalue of the symmetric rigidity matrix, here $\bar F_1(\mathbf p)$).
In view of the connections established in Section \ref{section: rigidity connection}, the techniques developed 
in these papers for estimating $\lambda_{\min}$ and maintaining it above a desired threshold 
are applicable, but a discussion of the resulting controllers is left for a full version of this paper.
In the next section, we focus on the computation of the gradient steps for the functions $f_D$ and $f_A$.

%


%% file: centralized.tex

%


\section{Gradient Computations for the Localizability Potentials}
\label{section: distributed algos}

To simplify the presentation, we assume in this section that the anchor robots are
fixed and compute the gradients only for the robots with unknown positions. Computing the
gradients of mobile anchors presents no additional difficulty.

\subsection{Partial Derivatives of the FIM}
\label{eq: FIM derivatives}

For $i=1,\ldots,n$, and $\nu = x$ or $y$, the partial derivatives of \eqref{eq: T optimal cost} read
%
\begin{equation}	\label{eq: fT gradient}
\frac{\partial f_T}{\partial \nu_i} (\mathbf p) = - \sum_{k=1}^n \Tr \left[ \frac{\partial F_{kk}}{\partial \nu_i} (\mathbf p) \right].
\end{equation}
For \eqref{eq: D optimal cost}, we deduce from \cite{minka2000old} that
\begin{equation}	\label{eq: fD gradient}
\frac{\partial f_{D}}{\partial \nu_i} (\mathbf p) = -Tr\left[ (F(\mathbf p))^{-1} \frac{\partial F}{\partial \nu_i}(\mathbf p) \right].
\end{equation}
Finally, for \eqref{eq: A optimal cost}, since $d f_{A} = Tr\left( d F^{-1} \right)$,
$d F^{-1} = (F^{-1}) \, dF \, (F^{-1})$ and $\Tr AB) = \Tr(BA)$, we have
\begin{equation}	\label{eq: fA gradient}
\frac{\partial f_{A}}{\partial \nu_i}(\mathbf p) = -Tr\left[ (F(\mathbf p))^{-2} \frac{\partial F}{\partial \nu_i}(\mathbf p) \right].
\end{equation}

In the equations above, we see that the expressions of $\frac{\partial F}{\partial x_i}(\mathbf p)$
and $\frac{\partial F}{\partial y_i}(\mathbf p)$ are needed.
%
Starting from \eqref{eq: non-diag block FIM}, \eqref{eq: diag block FIM}, we can obtain
the following expressions for the partial derivatives of the $F_{ij}$ blocks.
If $j \in \mathcal N_i$, then
\begin{align*}
&\frac{\partial F_{ij}}{\partial x_i} (\mathbf p_i, \mathbf p_j) = 
\frac{2}{\sigma^2 d_{ij}^{2 \alpha}} \times \\
&
\begin{bmatrix}
\alpha \frac{(x_i-x_j)^3}{d_{ij}^2} - (x_i-x_j)
& (y_i-y_j) \left(\alpha \frac{(x_i-x_j)^2}{d_{ij}^2} - \frac{1}{2} \right) \\ 
* & \alpha \frac{(x_i-x_j)(y_i-y_j)^2}{d_{ij}^2}
\end{bmatrix},
\end{align*}
where the symbol $*$ replaces the symmetric term, $\alpha$ is $1$ for additive noise and $2$ for multiplicative noise.
Similarly, 
\begin{align*}
&\frac{\partial F_{ij}}{\partial y_i} (\mathbf p_i, \mathbf p_j) = 
\frac{2}{\sigma^2 d_{ij}^{2 \alpha}} \times \\
&
\begin{bmatrix}
\alpha \frac{(y_i - y_j)(x_i-x_j)^2}{d_{ij}^2}
& (x_i-x_j) \left(\alpha \frac{(y_i-y_j)^2}{d_{ij}^2} - \frac{1}{2} \right) \\ 
* & \alpha \frac{(y_i-y_j)^3}{d_{ij}^2} - (y_i-y_j)
\end{bmatrix}.
\end{align*}

These expressions suffice to compute all the elements of $\frac{\partial F}{\partial x_i}(\mathbf p)$ and
$\frac{\partial F}{\partial y_i}(\mathbf p)$. For example, we have
$
\frac{\partial F_{ji}}{\partial x_i} (\mathbf p_j, \mathbf p_i) = \frac{\partial F_{ij}}{\partial x_i} (\mathbf p_i, \mathbf p_j)
$
by the symmetry of the functions $F_{ij}$. 
Then $\frac{\partial F_{ii}}{\partial x_i} (\mathbf p) = - \sum_{k \in \mathcal N_i} \frac{\partial F_{ik}}{\partial x_i} (\mathbf p_i, \mathbf p_k)$,
and for $k \neq i$, $\frac{\partial F_{kk}}{\partial x_i} (\mathbf p) = - \frac{\partial F_{ki}}{\partial x_i} (\mathbf p_k, \mathbf p_i)$.
Also, $\frac{\partial F_{kl}}{\partial x_i} (\mathbf p_k, \mathbf p_l) = 0$ if $k$ and $l$ are different from $i$,
or if $k$ and $l$ are not neighbors (since then $F_{kl} = 0$).
Decomposing the 
matrices $\frac{\partial F}{\partial x_i} (\mathbf p)$ and $\frac{\partial F}{\partial y_i} (\mathbf p)$ 
into $2 \times 2$ blocks, only the blocks $(i,i)$, the blocks $(i,j)$ and $(j,i)$ for $j \in \mathcal N_i$, and $(j,j)$ for $i \in \mathcal N_j$, are non zero.

We can then compute immediately the partial derivatives of $f_T$. Namely, starting from \eqref{eq: fT gradient},
\[
\frac{\partial f_T}{\partial x_i} (\mathbf p) = \Tr \left[ 
\sum_{k \in \mathcal N_i} \frac{\partial F_{ik}}{\partial x_i} (\mathbf p_i, \mathbf p_k) 
+ \sum_{k | i \in \mathcal N_k} \frac{\partial F_{ki}}{\partial x_i} (\mathbf p_k, \mathbf p_i)
 \right].
\]
Using our symmetric graph assumption and $F_{ik} = F_{ki}$,
\begin{align*}
\frac{\partial f_T}{\partial x_i}(\mathbf p) &= \frac{4}{\sigma^2} \sum_{k \in \mathcal N_i} \frac{1}{d_{ik}^{2 \alpha}}
\Big( -(x_i-x_k) \\
&\quad + \alpha \frac{(x_i-x_k)^3 + (x_i-x_k)(y_i-y_k)^2}{d_{ik}^2} \Big),
\end{align*}
and similarly
\begin{align*}
\frac{\partial f_T}{\partial y_i}(\mathbf p) &= \frac{4}{\sigma^2} \sum_{k \in \mathcal N_i} \frac{1}{d_{ik}^{2 \alpha}} 
\Big( -(y_i-y_k) \\
&\quad + \alpha \frac{(y_i-y_k)^3 + (y_i-y_k)(x_i-x_k)^2}{d_{ik}^2} \Big).
\end{align*}
We see that these expressions (for $\hat{\mathbf p})$ can be computed by agent $i$ by communicating only with its neighbors 
to obtain their relative position estimates $\hat{\mathbf p}_{ik} := \hat{\mathbf p}_i - \hat{\mathbf p}_k$.

%% file: distributed.tex
\subsection{Distributed Computations for D- and A-Optimal Design}
\label{eq: distributed algos}





For $f_D$ and $f_A$, the computations \eqref{eq: fD gradient} and \eqref{eq: fA gradient} of agent $i$ 
involve the diagonal blocks of $F^{-1} \partial F/\partial \nu_i$ and $F^{-2} \partial F/\partial \nu_i$, 
for $\nu = x$ or $y$.
Computing $\frac{\partial F}{\partial \nu_i}(\hat{\mathbf p})$ can be done by agent $i$ from only the knowledge of the
relative position estimates $\hat{\mathbf p}_{ij}$ with respect to its neighbors in the ranging graph, obtained from 
the cooperative localization scheme.
%
%
In the following discussion, we refer to the decomposition of the matrices $F$ and $\partial F/\partial \nu_i$ 
in terms of $2 \times 2$ blocks as in Sections \ref{eq: FIM expressions} and \ref{eq: FIM derivatives}.
%
Now recall that the columns $k$ of $\frac{\partial F}{\partial \nu_i}$ are nonzero only for 
$k \in \mathcal N_i$ and $k=i$. Denoting these columns by $b_k$, we get
\begin{align} \label{eq: gradient formula}
\frac{\partial f_D}{\partial \nu_i}(\mathbf p) = -\Tr \left( [F^{-1} b_i]_i \right) - \sum_{j \in \mathcal N_i} \Tr \left( [F^{-1} b_j]_j \right), 
\end{align}
where $[\cdot]_k$ means the two rows corresponding to agent $i$ (so $[F^{-1} b_k]_l$ is a $2 \times 2$ matrix). 

We have then reduced the problem to computing $[F^{-1} b_k]_k$ in a distributed manner, for $k = i$
and $k \in \mathcal N_i$. For this, consider, at the current (fixed) estimate $\hat{\mathbf p}$, the following 
continuous-time system
\begin{equation}	\label{eq: distributed updates}
\dot \xi = - k(F(\hat{\mathbf p}) \, \xi - c),
\end{equation}
where $k > 0$ and $\xi(t), c$ are $2n \times 2$ matrices.
Because $F$ is positive definite, hence with positive eigenvalues,
this system converges to a steady-state $\xi_\infty$ satisfying $F \xi_\infty = c$, i.e., $\xi_\infty = F^{-1} c$.
Moreover, because of the structure of $F$, which is that of a Laplacian matrix, the updates
\eqref{eq: distributed updates} are inherently distributed, with agent $k$ updating
the component $\xi_{kk}$ (a $2 \times 2$ matrix) by communicating only with its neighbors.

We can use the scheme \eqref{eq: distributed updates} to compute $F^{-1} b_i$ by taking $c = b_i$. 
Note that each agent $k$ knows its $2 \times 2$ matrix $[b_{i}]_k$, which is either zero if $k \notin \mathcal N_i$,
or requires only the knowledge of the relative state with respect to $i$ if $k \in \mathcal N_i$, and the expressions
of Section \ref{eq: FIM derivatives}.
After convergence, agent $i$ knows $[F^{-1} b_i]_i$. 
In parallel, we do the same to compute  $F^{-1} b_j$ for $j \in \mathcal N_i$,
and after convergence, agent $j$ knows $[F^{-1} b_j]_j$. Finally, the agents $j \in \mathcal N_i$ can send
their $2 \times 2$ matrices to $i$ in order to compute the gradient \eqref{eq: gradient formula}.
The iterations corresponding to \eqref{eq: distributed updates} can be stopped when only rough
convergence is obtained, i.e., the gradient direction only approximately computed,
since the goal of the algorithm is to minimize $f_D$ and not to compute the gradient at each step 
very accurately. 
Indeed, this idea is the foundation of algorithms such as stochastic gradient algorithms.


Finally, to compute the gradients of $f_A$, one possible approach is to allow robots
to exchange information with their two-hop neighbors. The discussion above 
can then be repeated with $F^2$ replacing $F$.

%% file: simulations.tex

\section{Summary and Simulations}
\label{section: simu}

The proposed approach to localizability-constrained motion planning can be summarized as follows.
Initially, the robots are in some configuration $\mathbf p^0$, forming a ranging network that we assume 
to be sufficiently connected for $F(\mathbf p^0)$ to be positive definite. In particular, this requires enough
links with the anchor nodes, whose positions are known. The algorithm then proceeds in successive steps.
At step $k$, the robots run the cooperative localization algorithm and each robot knows an estimate $\hat{\mathbf p}_i^k$
of its position. The robots then follow a gradient step \eqref{eq: individual gradient step}. Computing
this step involves for the localizability potential the distributed iterations described in 
Section \ref{eq: distributed algos}, which must thus be executed on a faster time scale than the motion
so that $\hat{\mathbf p}_k$ can be considered constant during these iterations.

\begin{figure}
\centering
\includegraphics[width=0.8\linewidth]{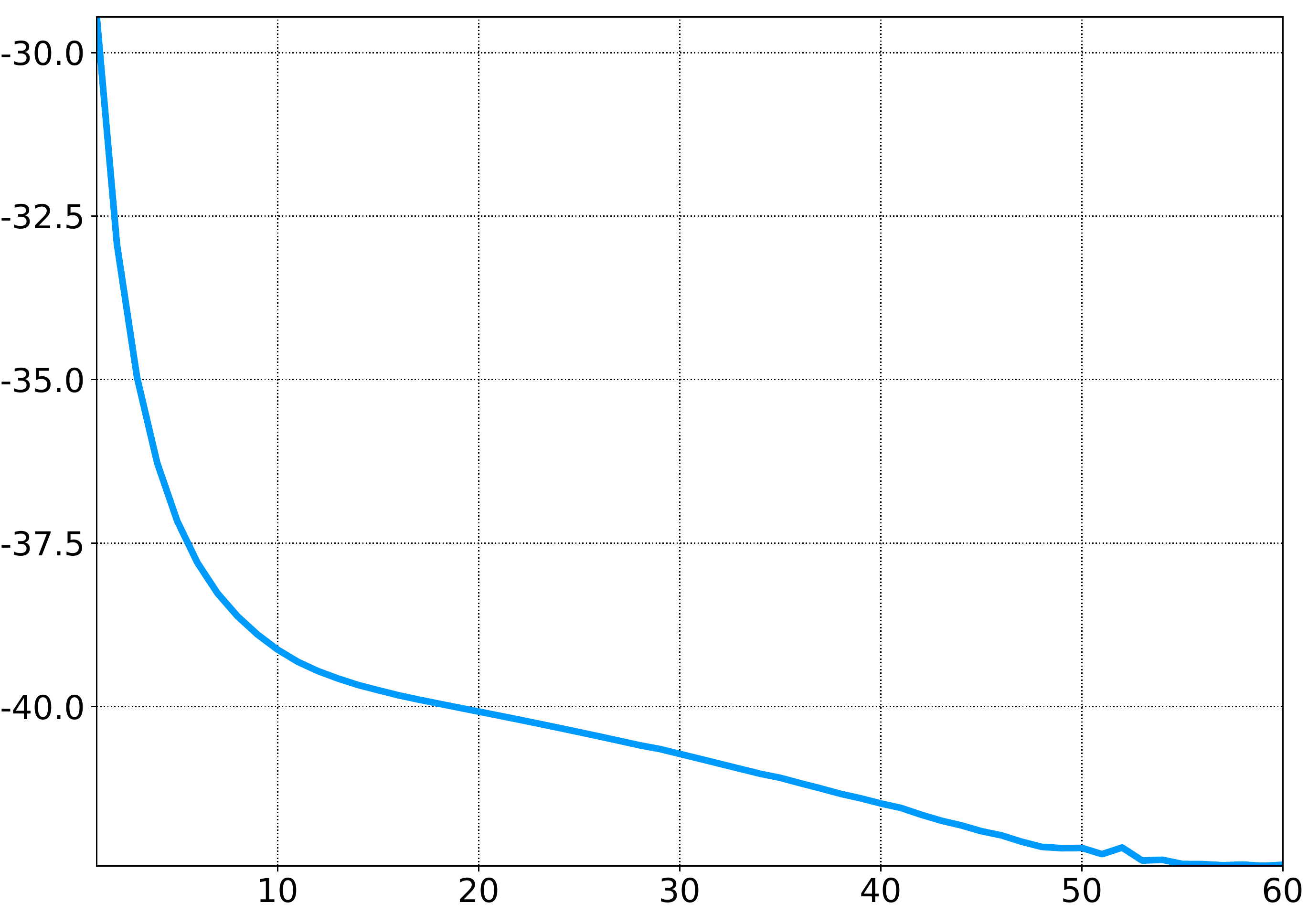}
\caption{Cost $f_D(\mathbf p^k)$ as a function of $k$.}
\label{fig: cost fn}
\end{figure}

\begin{figure}
\centering
\includegraphics[width=0.8\linewidth]{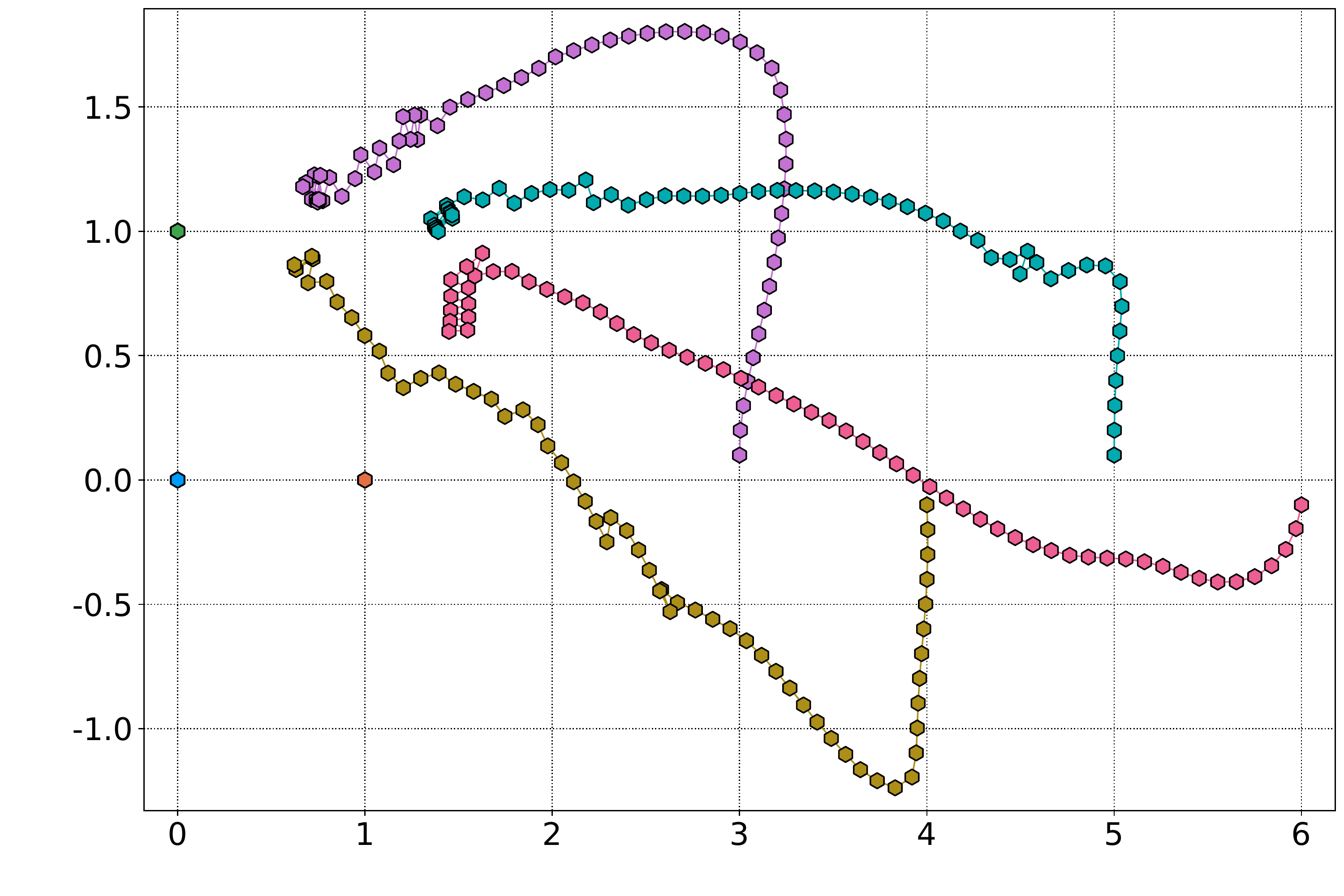}
\caption{Robot trajectories descending the gradient of $f_D$.}
\label{fig: trajs}
\end{figure}

\begin{figure}
\centering
\includegraphics[width=0.8\linewidth]{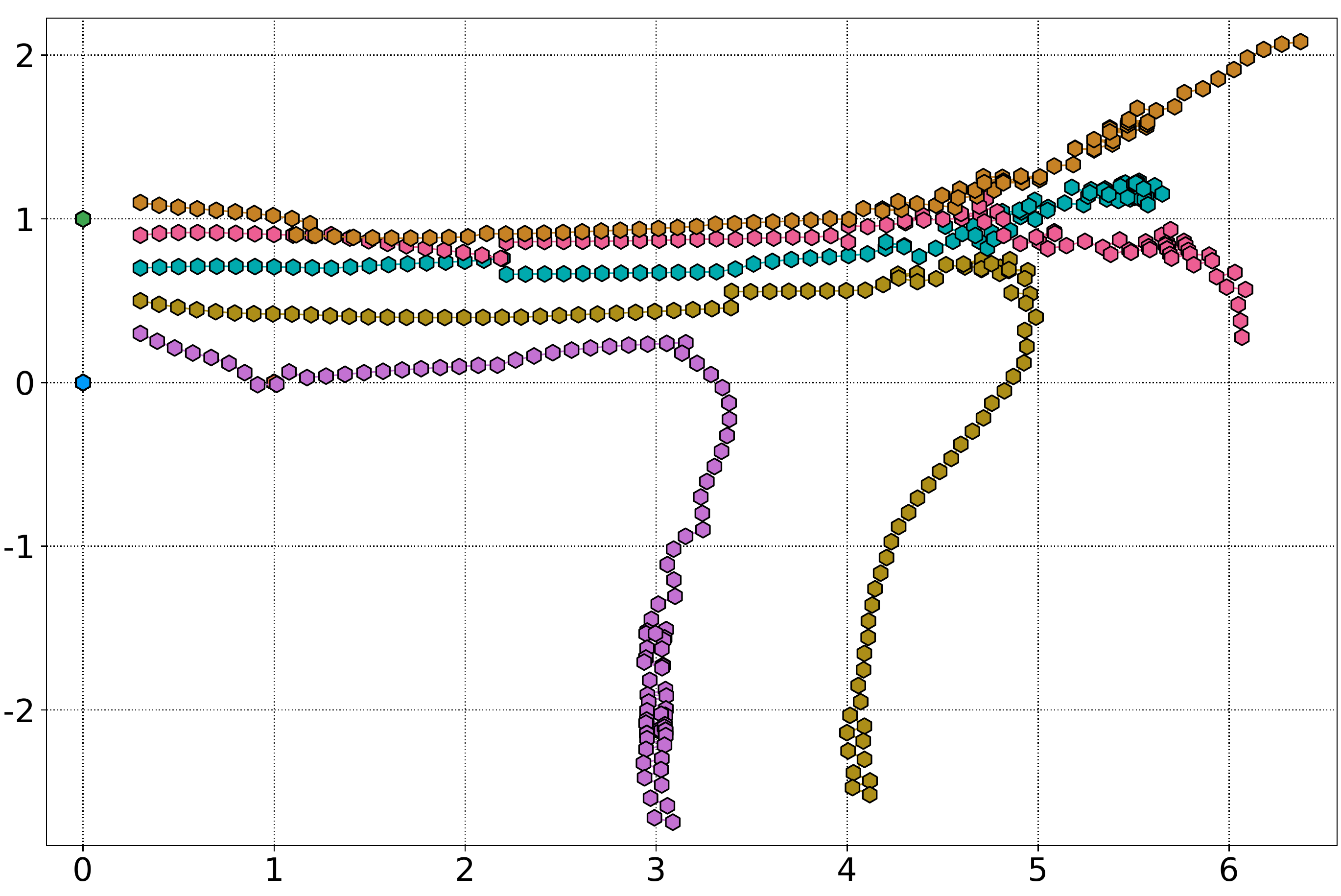}
\caption{Trajectories for the full potential \eqref{eq: global potential}.}
\label{fig: trajs full}
\end{figure}

For illustration purposes, we show on Fig. \ref{fig: cost fn} and \ref{fig: trajs} the evolution of the 
localizability potential $f_D$ and the trajectories of the $4$ robots descending the gradient of $f_D$. 
We assume a perfect position estimator in order to keep the discussion focused on the motion planning component.
There are $3$ static anchors at coordinates $[0,0]$ (anchor 1), $[1,0]$ (anchor 2) and $[0,1]$ (anchor 3).
The mobile robots start from the vicinity of $[i,0]$ for $i=3,\ldots,6$, so they are initially almost aligned.
All links in the ranging graph are present except for the robots $6$ and $7$ starting close to
$x=5$ and $x=6$, which do not measure the distances to the anchors $1$ and $3$.
We assume additive ranging noise ($\alpha = 1$) with $\sigma = 0.1$. 
We see that the robots eventually move back toward the anchors, but initially they try 
to break their alignment by moving away from the $x$-axis, in order 
to reach a better geometry. This is reflected in the initial sharp drop in the potential.
Finally, Fig. \ref{fig: trajs full} shows the trajectories of $5$ mobile robots starting from the neighborhood 
of the same anchors, for the full potential function \eqref{eq: gradient descent - estimate}, with
$f_{loc}=f_D$ and targets $x$-coordinates $\bar x_i$ in \eqref{eq: task potential} equal to $3, 4, 5, 6$ and $7$.

